\documentclass[%
reprint,
showkeys,
amsmath,amssymb,
aps]{revtex4-2}

\usepackage{dcolumn}
\usepackage{bm}
\usepackage{multirow}
\usepackage{graphicx}
\usepackage{hyperref}
\usepackage{subfigure}
\usepackage[dvipsnames]{xcolor}
\newcommand{\ket}[1]{\left| #1 \right\rangle}

\begin{document}
	
	\preprint{APS/123-QED}
	
	\title{Ultracold Interactions between Ions and Polar Molecules}
	
	\author{Leon Karpa}
	\affiliation{Leibniz Universität Hannover, Institut für Quantenoptik, 30167 Hannover, Germany.}
	\email{karpa@iqo.uni-hannover.de}
	\author{Olivier Dulieu}
	\affiliation{Universit\'e Paris-Saclay, CNRS, Laboratoire Aim\'e Cotton, Orsay 91400, France.}
	
	\date{\today}
	
	\begin{abstract}
		We propose a platform for observing and controlling the interactions between atomic ions and a quantum gas of polar molecules in the ultracold regime. This approach is based on the combination of several recently developed methods in two so-far complementary research domains: ion-atom collisions and studies of ultracold polar molecules. In contrast to collisions between ions and ground-state atoms, which are dominated by losses due to three-body recombination (TBR) already at densities far below those typical for quantum degenerate ensembles, our proposal makes use of polar molecules, their rich level structure, and sensitivity to electric fields to design effective interaction potentials where ion-neutral TBR losses and molecule-molecule losses due to sticky collisions could be strongly suppressed. This may open a broad range of applications including precise control of collisional properties in molecular ensembles using ions, quantum simulations, and cold quantum chemistry between polyatomic molecules.
	\end{abstract}
	
	\keywords{ion-atom interactions, molecular quantum gases, quantum chemistry}
	\maketitle
	
	\section{Introduction}
	Controlling collisions between ions and neutral atoms in a regime where quantum effects are dominant is a long-standing goal in the field of ion-atom interactions \cite{Tomza2019,Lous2022}. This comparatively new and interdisciplinary research domain holds great promise in advancing or enabling a broad spectrum of applications in areas including quantum many-body and polaron physics \cite{Goold2010,Schurer2017,Astrakharchik2023,Cote2002,Christensen2022,Chowdhury2024}, charge mobility \cite{Cote2000,Dieterle2021}, quantum simulations (QS) \cite{Bissbort2013,Gerritsma2012}, quantum chemistry (QC) \cite{Smith2005,Grier2009,Zipkes2010,Willitsch2008,Rellergert2011,Li2020,Weckesser2021}, and quantum information processing (QIP) \cite{Doerk2010,Secker2016,Ebgha2019}. Most of these applications either strongly benefit from or require ultralow temperatures where collisional processes are dominated by the contribution of the lowest partial wave, the so-called $s$-wave regime. During the last few years, enormous progress has been made towards this goal. In particular, recent advances have enabled several key achievements such as the elimination of micromotion-induced heating in generic compound optical traps \cite{Schmidt2020}, sympathetic cooling below the Doppler limit and close to the $s$-wave regime \cite{Feldker2020,Weckesser2021,Schmidt2020}, and the controlled association of ion-atom Feshbach molecules \cite{Weckesser2021}.
	
	However, some substantial challenges remain an obstacle in view of many envisioned applications which require access to the deep quantum regime characterized by a combination of quantum degenerate atomic gases and ions in or close to the motional ground state. This is largely due to inelastic or reactive collisions which dominate already at moderate neutral densities \cite{Haerter2012,Schmidt2020,Mohammadi2021} even for spin-polarized fermionic ensembles \cite{Feldker2020,Weckesser2021} and lead to loss of the ion. They originate from the comparatively large characteristic range $R^{\star}$ in the typical order of $100 ~ \textrm{nm}$ between an electric charge and the induced dipole moment of the atoms exposed to the electric field of the ion. For ground-state atoms located at the distance $R$ from the ion, this attractive interaction potential scales as $1/R^4$ and manifests itself in the form of strongly enhanced ion-atom-atom three-body recombination processes (TBR) \cite{Haerter2012,Krukow2016,Dieterle2020,Pandey2024}. At the same time, the attractive interaction is also responsible for a displacement of the ion from the node of the radiofrequency (rf) quadrupole field used to generate the confining potential for the ion, facilitating energy transfer from the rf field to the center of mass motion of the ion-atom system whenever they are allowed to undergo collisions \cite{Cetina2012}.
	
	This so-called micromotion-induced heating has been identified as one of the principal obstacles to reaching the $s$-wave regime in conventional hybrid traps utilizing rf fields to confine the ions \cite{Cetina2012}. In order to circumvent these limitations, several strategies have been brought forward making use of different systems and mechanisms such as Rydberg atom-atom interactions \cite{Kleinbach2018,Zuber2022}, Rydberg dressing \cite{Secker2017}, restricting the studies to few specific combinations with an extremely low reduced mass \cite{Cetina2012,Feldker2020,Weckesser2021}, and bichromatic traps for confining ions and atoms \cite{Schmidt2020,Karpa2021}. While all these strategies have enabled recent breakthroughs, the combination of ions with quantum degenerate atomic ensembles which are characterized by large phase-space densities on the order of 1 and high densities $n$ typically of the order of $10^{15} ~ \textrm{cm}^{-3}$, is still out of reach. This is a consequence of the strong enhancement of TBR in vicinity of an ion in conjunction with the scaling of the related loss rates as $n^2$ resulting in drastically reduced ion lifetimes expected for the required densities \cite{Haerter2012,Schmidt2020,Weckesser2021}.
	
	In parallel to the rapid evolution of studies focusing on ion-atom combinations, the complementary field of molecular quantum gases has seen a similarly impressive development. While the focus of the field shifted from chemically reactive bialkali mixtures investigated in first ground-breaking experiments \cite{Ospelkaus2010,Ni2008} to nonreactive combinations and directly laser-cooled molecular species, the surprisingly limited lifetimes observed for all species emerged as one of the main challenges \cite{Mayle2013,Croft2023,Jachymski2022,Gersema2021,Bause2021,Gregory2020,Hu2019}. For non-reactive combinations, the underlying losses have prevented the implementation of evaporative cooling, and until recently, hampered the creation of quantum degenerate gases. The exact mechanism behind these losses remains unclear and is a subject of ongoing debates \cite{Croft2023,Jachymski2022}, but it is likely related to so-called sticky collisions, which occur at short intermolecular distances and involve the formation of metastable collisional complexes  \cite{Mayle2013,Christianen2019,Croft2023,Jachymski2022}. With the development of methods to prevent the molecules from reaching the short range \cite{Gorshkov2008,Quemener2011,Lassabliere2018,Karman2018,Karman2019,Xie2020,Karam2023}, this obstacle has recently been overcome \cite{Matsuda2020,Anderegg2021}. Such shielding methods rely on engineered potential barriers created through interactions with microwave \cite{Lassabliere2018,Karman2018,Anderegg2021}, optical \cite{Karam2023,Xie2020} or static electric fields \cite{Quemener2010,Matsuda2020}. They have enabled efficient evaporative cooling \cite{Schindewolf2022,Lin2023,Bigagli2024}, and the first observations of molecular Fermi gases \cite{DeMarco2019,Schindewolf2022} and Bose-Einstein Condensates (BECs) \cite{Bigagli2024} have now become a reality. %
	Similarly, a series of breakthrough achievements in the effort to control polar molecules with ever greater precision, now enables experiments on the level of individual particles, e.g., single molecules in optical tweezers \cite{Zhang2022a,Gregory2024,Vilas2024,Holland2023}, as well as directly laser-coolable \cite{VazquezCarson2022, Holland2023, Langin2021, Zeng2024, Burau2024}, and even polyatomic molecules \cite{Vilas2024, Mitra2020}. 
	
	Here, we discuss the possibility to combine the methodology and advantageous properties of the respective systems employed in both fields in order to extend experimental studies of ion-neutral interactions at ultralow temperatures towards quantum gases of polar molecules. This platform would provide additional degrees of freedom enabling minute control over several simultaneously active interactions. We show that by exploiting these properties, characteristic quantities such as $R^{\star}$ and rates for reactive collisions could be tuned by orders of magnitude, allowing for a broad spectrum of new applications. Our article is structured in the following way: in Section \ref{molecules_approaching_ions}, we derive characteristic quantities of collisions between ions and polar molecules considering the influence of internal rotational degrees of freedom. We identify combinations and specific rotational states suitable for studies at ultralow collision energies. Subsequently, we discuss how existing methodology can be used to realize experimental platforms capable of combining ions and polar molecules, exemplified for Ba$^{+}$-NaK, in Section \ref{toolbox}. In the rest of the paper, we consider two regimes of neutral molecule density, that is, dilute molecular gases suitable for novel studies of quantum chemistry (Section \ref{reactive collisions}), and dense quantum gases where the interplay of attractive and repulsive long-range and short-range interactions is expected to play an important role (Section \ref{densemolecules}). Lastly, we discuss potential applications of the proposed hybrid ion-molecule system in studies of quantum-chemical reactions, many-body phenomena, and dynamical processes in molecular quantum gases as well as their control in Section \ref{applications}.
	
	\section{Polar molecules in the electric field of an atomic ion} \label{molecules_approaching_ions}
	
	\subsection{Molecules in the rovibronic ground state}
	\begin{figure}[b]
		\begin{center}
			\includegraphics[width= 0.5 \textwidth]{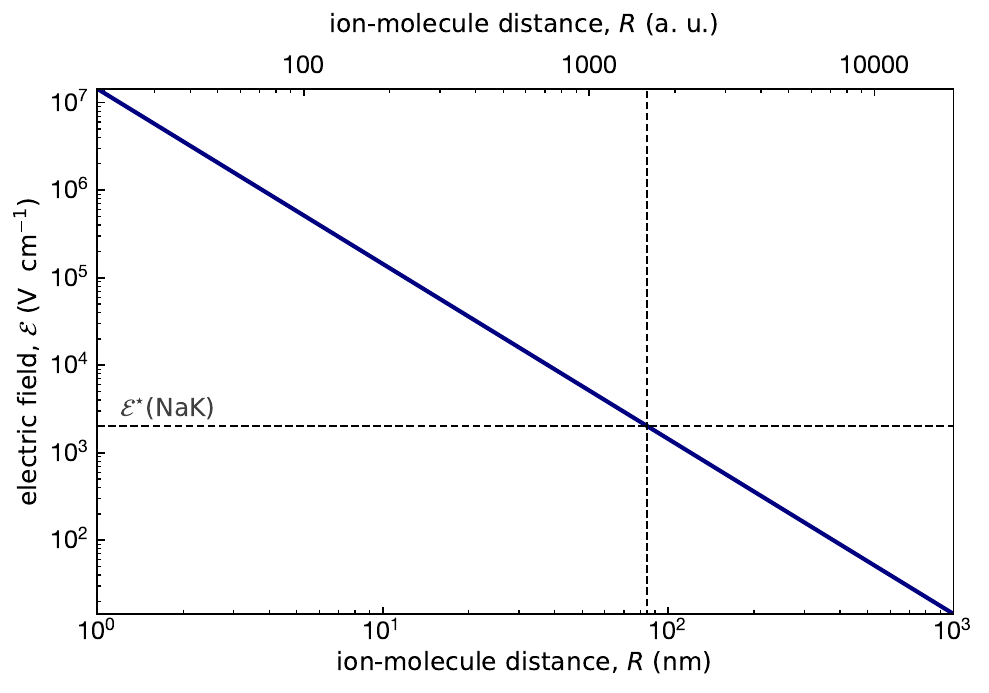}\\
			\includegraphics[width= 0.495 \textwidth]{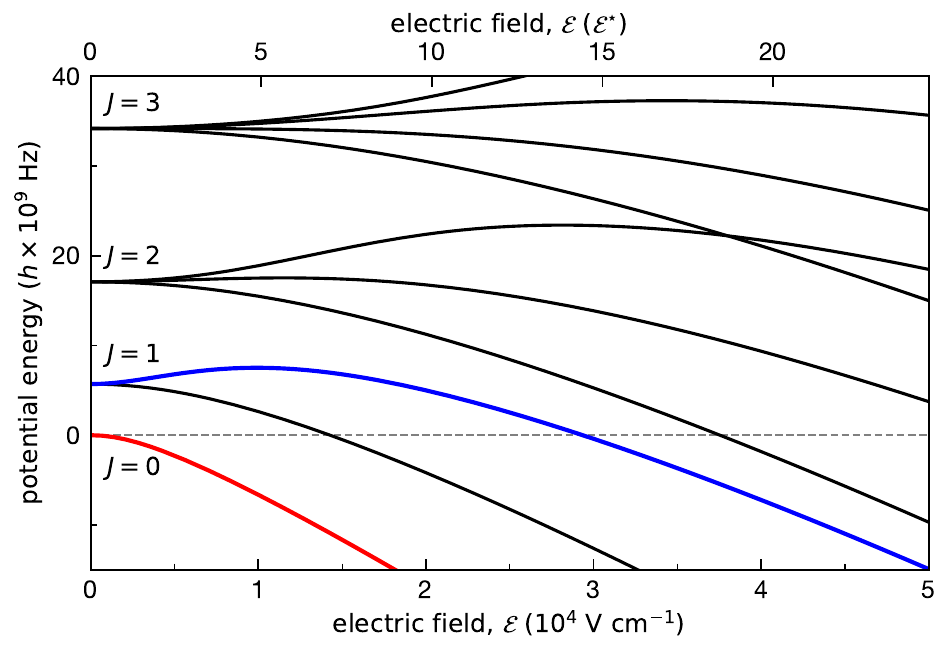}
		\end{center}
		\caption{
			\textbf{(top panel)} Magnitude of the electric field emanating from a single ion as a function of its distance to an approaching neutral molecule. The gray dashed line denotes the characteristic electric field of NaK, \linebreak $\mathcal{E}^{\star}$(NaK)$= B_0/d_0 = 2.04 ~ \textrm{kV cm}^{-1}$. \textbf{(bottom panel)} Numerically calculated energy levels of the NaK molecule for the lowest rotational states $\ket{J=0, ~1, ~2, ~3; m_J}$ (considering a maximal $J_{max}=10$) within the vibronic ground state manifold $\ket{X^{1}\Sigma^{+}, v=0}$ as a function of an external electric field of magnitude $\mathcal{E}$. The hyperfine structure which is on the order of $h \times 100~\textrm{kHz}$ is neglected for simplicity. The lowest high and low-field seeking states (up to $\mathcal{E}$ $\approx 5 ~ \mathcal{E}^{\star}$), $ \ket{J = 0, m_J=0} $ and $ \ket{J = 1, m_J=0} $, are shown as red and blue solid lines, respectively.}
		\label{fig:eigenenergies}
	\end{figure}
	
\begin{figure}[!htb]
		\begin{center}
			\includegraphics[width= 0.5 \textwidth]{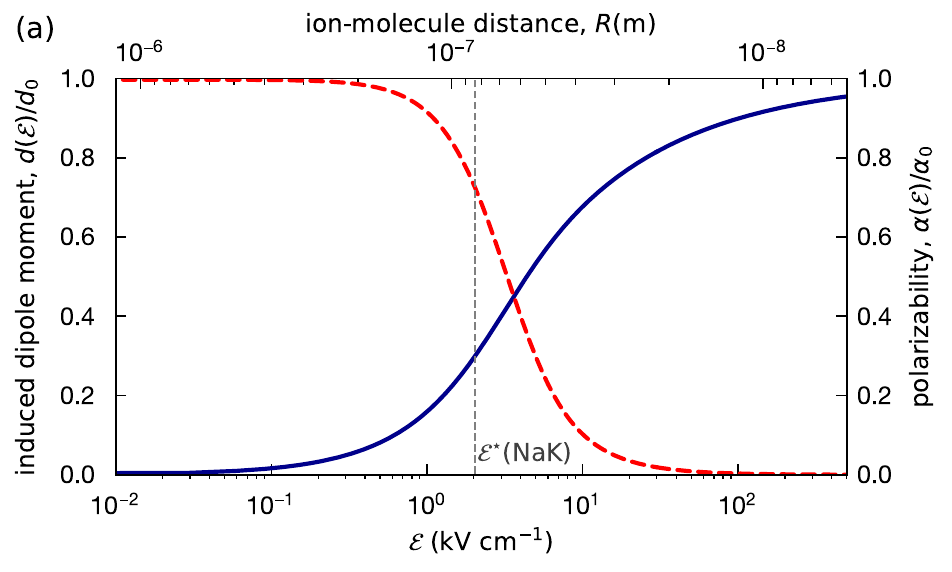}
			\includegraphics[width= 0.5 \textwidth]{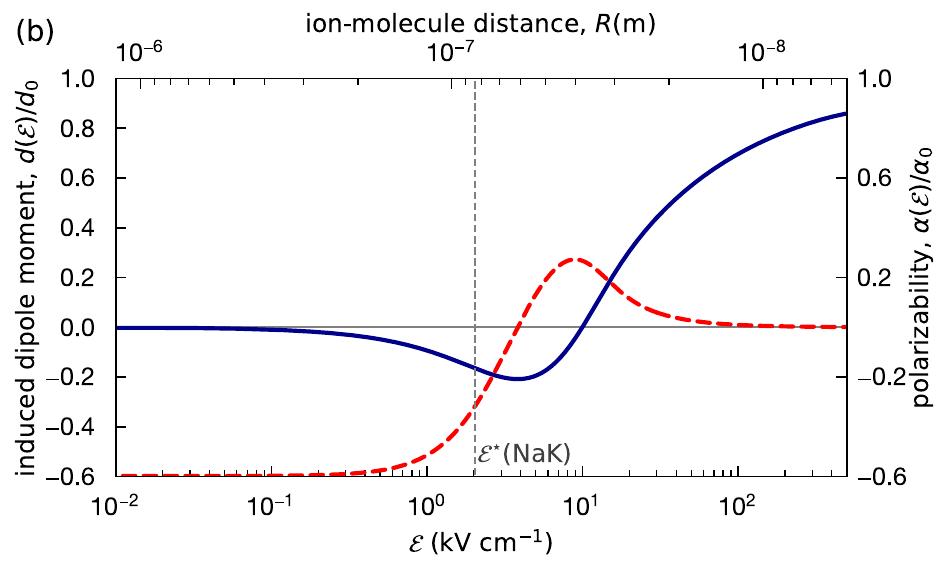}
		\end{center}
		\caption{Induced dipole moment $ d(\mathcal{E})$ (blue solid line) and the static polarizability $\alpha( \mathcal{E}) $ (red dashed line) for NaK as a function of the electric field $ \mathcal{E}$ created by an ion separated by the distance $R$ from the (stationary) molecule. \textbf{(a)} Calculated $d(\mathcal{E})$ in units of the permanent dipole moment $d_0$ and $\alpha( \mathcal{E}) $ in units of $\alpha_0$ (taken fom \cite{Vexiau2017}) for the rovibronic ground state $\ket{X^{1}\Sigma^{+}, v=0, J = 0, m_J=0}$ and \textbf{(b)} the $\ket{X^{1}\Sigma^{+}, v=0, J = 1, m_J=0}$ state. The characteristic electric field $\mathcal{E}^{\star} $ is shown as the horizontal dashed line.}
		\label{fig:polarization}
	\end{figure}

	First, we consider the situation where an ion is approached by a single polar molecule prepared in its rovibronic ground state $\ket{X^{1}\Sigma^{+}, v=0, J = 0, m_J=0}$ at a given distance $R$, with $v$, $J$ and $m_J$ denoting the vibrational and rotational quantum number and the projection of $\vec{J}$ onto the electric field $\vec{\mathcal{E}}$, respectively.  This situation can be treated in the same way as the interaction of induced dipoles in a radially symmetric external electric field created by a pointlike charge \footnote{The corresponding dipole approximation is valid for ion-molecule separations that are large compared to the bond length of the molecule which is the case in all scenarios discussed in this work. A more general description requires a multipole expansion of the ion-neutral interaction, e.g., in systems such as Rydberg atoms in proximity of ions
		\cite{Duspayev2021, Deiss2021, Zuber2022}.}, as depicted in Fig.~\ref{fig:eigenenergies} (top panel). The charge-dipole interaction of each individual molecule with the ion is determined by the Stark effect \cite{Aldegunde2008}:
	\begin{equation}
		\hat{H}_{\textrm{CD}} = -  \vec{d}( \vec{\mathcal{E}}) \cdot \vec{\mathcal{E}}(R).
		\label{eqn:CDinteraction}
	\end{equation}
	The competition between the molecule's rotation and the alignment of its dipole moment $\vec{d}( \vec{\mathcal{E}})$ can be expressed in terms of the characteristic electric field $\mathcal{E}^{\star}  = B_0 / d_0$, where $ B_0$ is the rotational constant and $d_0$ the permanent dipole moment (PDM) of the molecule in the body-fixed frame. The rotational energy in the vibrational ground state $v=0$ is approximately $U_{\textrm{rot}} \approx B_0 ~ J( J +1)$. For large ion-molecule separations corresponding to the low-field limit where $\mathcal{E} \ll \mathcal{E}^{\star}$ the eigenvalues $U_{\textrm{CD}}(\mathcal{E})$ of $\hat{H}_{\textrm{CD}}$ are quadratic in $\mathcal{E}$:
	\begin{equation}
		U_{\textrm{CD}}(\mathcal{E}) = U_{\textrm{rot}} - \left.{ d (\mathcal{E}) }\right\vert _{\mathcal{E}=0} \mathcal{E} - \frac{1}{2} \left.{\alpha(\mathcal{E}) } \right \vert _{\mathcal{E}=0} \mathcal{E}^2 + O( \mathcal{E}^3),
		\label{eqn:ESeries}
	\end{equation}
	with the induced dipole moment and static polarizability being:
	\begin{equation}
		d (\mathcal{E}) = - \frac{\partial U_{\textrm{CD}}(\mathcal{E})}{\partial \mathcal{E}}, ~
		\alpha (\mathcal{E}) = - \frac{\partial^2 U_{\textrm{CD}}(\mathcal{E})}{\partial \mathcal{E}^2}.
		\label{eqn:derivatives}
	\end{equation}
	As an example, for the eigenvalues of the rovibronic ground state of NaK from Fig.~\ref{fig:eigenenergies} (bottom panel, red solid line), these quantities are shown in Fig.~\ref{fig:polarization}.
	Utilizing the ion with an electric charge $q$ as a point-like source of an external electric field $\mathcal{E} = q/(4 \pi \epsilon_0 R^2)$ for the surrounding molecules, we obtain the long-range ion-molecule potential
	\begin{equation}
		U_{\textrm{CD}}(R) | _{R\rightarrow \infty} = - \frac{1}{2} \alpha_0 \mathcal{E}^2 = - \frac{1}{2} \frac{\alpha_0 q^2}{(4 \pi \epsilon_0)^2} \frac{1}{ R^4},
		\label{eqn:CIDinteraction}
	\end{equation} 
	with the vacuum permittivity $\epsilon_0$. This result is identical with the familiar expression for charge-induced dipole interactions which dominate ion-atom collisions where the long-range induction coefficient is given by \linebreak $C_4 = \frac{1}{2}\alpha_0 q^2/(4 \pi \epsilon_0)^2$  \cite{Cote2000a}. This allows us to calculate $C_4$ coefficients for polar molecules and to directly compare the expected long-range behavior with the well-explored ion-atom systems. Analogously, we use the following expressions derived from the semiclassical Langevin scattering model \cite{Langevin1905,Metcalf1999}, to calculate the range of this potential and the characteristic energy scale for $s$-wave collisions:
	\begin{equation}
		R^{\star} = \sqrt{2 \mu C_4 / \hbar^2 }
	\label{eqn:Rstar}
	\end{equation}
	\begin{equation}
		E^{\star} = \frac{\hbar^4}{4 \mu^2 C_4} = \frac{\hbar^2}{2 \mu (R^{\star})^2}.
	\label{eqn:Estar}
	\end{equation}
	Here, $\hbar$ and $\mu$ denote the reduced Planck's constant and the reduced mass of the ion-neutral system, respectively.
	\begin{table*}
		\caption{\label{tabone} Characteristic properties of the charge-induced dipole long-range interaction between common heteronuclear bialkali and laser-coolable molecules prepared in the rovibronic ground state and a $^{138}$Ba$^{+}$ ion. The static dipole polarizabilities $\alpha_0$ and permanent electric dipole moments (PDM) of the bialkali molecules are taken from \cite{Vexiau2017}. The polarizabilities due to purely rotational transitions of the laser-coolable species were approximated  as $\alpha_0^{\textrm{rot}} \approx d_0^2/(3 B_0)$ \cite{Lepers2013} with constants $d_0$ and $B_0$ for CaH \cite{Steimle2004,Huber1979}, CaF  \cite{Childs1984}, and CaOH \cite{Steimle1992, Fletcher1995}. For comparison, $^{87}$Rb and $^{6}$Li are given as references for Ba$^{+}$-atom systems, representing a typical and one of the most favorable combinations with respect to reaching the $s$-wave scattering limit, respectively.
		} 
		\begin{ruledtabular}
			\begin{tabular} {rddddd}
				&    \multicolumn{1}{c}{PDM} &  \multicolumn{1}{c}{$B_0/h$} & \multicolumn{1}{c}{$\alpha_0$} &   \multicolumn{1}{c}{$R^{\star}$} &    \multicolumn{1}{c}{$E^{\star}/k_{\textrm{B}}$} \cr 
				&   \multicolumn{1}{c}{(D)} & \multicolumn{1}{c}{(GHz)} &  \multicolumn{1}{c}{$(10^{-37}~ \textrm{C m}^2 \textrm{V}^{-1})$} &    \multicolumn{1}{c}{$( 10^{-6}  \textrm{m} ) $} &   \multicolumn{1}{c}{(nK)} \cr
				\colrule
				LiCs & 5.59 & 5.61 & 312 & 25.9 & 0.00521 \\
				NaCs & 4.69 & 1.74 & 709 & 40.1 & 0.00206 \\
				LiRb & 4.18 & 6.45 & 152 & 16.2 & 0.0165 \\
				LiK & 3.58 & 7.67 & 93.6 & 10.0 & 0.0703 \\
				NaRb & 3.31 & 2.10 & 294 & 23.6 & 0.00711 \\
				NaK & 2.78 & 2.85 & 152 & 14.2 & 0.0281 \\
				KCs & 1.84 & 0.899 & 208 & 22.2 & 0.00644 \\
				RbCs & 1.25 & 0.48 & 177 & 21.6 & 0.00614 \\
				KRb & 0.62 & 1.14 & 18.8 & 6.21 & 0.0952 \\
				LiNa & 0.57 & 11.2 & 1.64 & 1.12 & 7.86 \\
				\colrule
				CaH & 2.94 & 128 & 3.77 & 1.92 & 2.08 \\
				CaF & 3.07 & 10.3 & 51.4 & 8.11 & 0.0893 \\
				CaOH & 1.47 & 10.0 & 51.4 & 8.01 & 0.0937 \\
				\colrule
				Rb & 0.00 & 0.00 & 0.0526 & 0.295 & 52.3 \\
				Li & 0.00 & 0.00 & 0.0270 & 0.0694 & 8760 \\
			\end{tabular}
		\end{ruledtabular}	
		\label{tab:moleculeoverview}
	\end{table*}

	When considering the rovibronic ground state, one of the most significant features of heteronuclear molecules is that the magnitude of the static polarizability is typically several orders of magnitude larger than that of atoms as summarized in Table \ref{tab:moleculeoverview} for the case of an immersed Ba$^{+}$ ion. As pointed out in \cite{Vexiau2017}, this enormous enhancement comes about because $\alpha_0$ varies as the inverse of the energy for the lowest allowed transition from the rovibronic ground state. While the dipole moment of transitions between rotational states is comparable to that of dipole allowed transitions to electronically excited states, the energy separation of rotational states is typically of the order $B_0/h = 1\text{--}10 ~ \textrm{GHz}$. The latter is about five orders of magnitude lower than the energy of typical electronic transitions, such that $\alpha_0$ and the $C_4$ coefficients are correspondingly larger. For instance, the combination Ba$^{+}$ and NaK, with the latter being prepared in the rovibronic ground state and possessing a PDM of $d_0 = 2.78~ \textrm{D}$ as well as a static electric dipole polarizablility of \linebreak $\alpha_0 = 9.237 \times 10^{5} ~ \textrm{a.u.}$ \cite{Vexiau2017}, yields the long-range induction coefficient $C_4^{\textrm{Ba-NaK}} = 1.58 \times 10^{-53} ~ \textrm{C V } \textrm{m}^{4} $ which is approximately $3 \times 10^{3} ~ C_4^{\textrm{Ba-Rb}}$. For the resulting characteristic range of the interaction and the related energy scale we obtain $R^{\star}_{\textrm{NaK}} \approx 14 ~ \mu\textrm{m}$ and $E^{\star}_{\textrm{NaK}}/k_{\textrm{B}} \approx 3 \times 10^{-11} ~\textrm{K}$, respectively. This straightforward result already reveals two principal properties of the expected ion-molecule interactions. Firstly, the range of the charge-induced dipole potential is extremely large compared to the characteristic range of ion-atom interactions which is typically on the order of $100~ \textrm{nm}$, e.g., $R^{\star}_{\textrm{Rb}} \approx 300 ~ \textrm{nm}$ for Rb prepared in its electronic ground state. Secondly, compared to ion-atom mixtures, the related energy threshold for entering the $s$-wave regime is drastically reduced even further by several orders of magnitude. For comparison, the $s$-wave threshold in rubidium is $E^{\star}_{\textrm{Rb}}/k_{\textrm{B}} \sim 10^{-8} ~\textrm{K}$ while the most favorable ion-atom mixtures such as Yb$^{+}$-Li and Ba$^{+}$-Li feature a much higher threshold on the order of $k_{\textrm{B}} \times 10^{-6} ~\textrm{K}$ \cite{Cetina2012}.
	
	In addition to the seeming difficulty of reaching the required extremely low collision energies, another effect potentially poses an even stricter limitation on the range of observable effects: the enhancement of the loss processes due to three-body recombination. From ion-atom collisions it is known that ion-atom-atom TBR becomes the dominant loss process already at comparatively low densities several orders of magnitude below values where atomic ensembles suffer from three-body losses \cite{Haerter2012}, even in the case of spin-polarized fermions \cite{Weckesser2021}. This can be understood intuitively by coarsely estimating the critical density $n_{\textrm{C}}$ at which the interparticle volume becomes smaller than the active volume %
	defined here as:%
	\begin{equation}
		V^{\star} = \frac{4}{3} \pi (R^{\star}) ^3. 
	\end{equation}
	 The latter is determined by the characteristic range of the dominant interaction $-C_p/R^p$ \cite{Metcalf1999}:
	 \begin{equation}
	 	R^{\star}(p) = \left (  \frac{p \mu C_p}{2 \hbar^2}  \right)^{1/(p-2)}.
	 \end{equation}
	 For $n>n_{\textrm{C}}= (V^{\star})^{-1}$ more than one particle enter $V^{\star}$ at the same time. At this point TBR becomes the dominant loss process according to the simplified model, while under realistic conditions, TBR losses are significant already at lower densities. In the case of ground-state neutral atoms, $R^{\star}$ arises from the short-range van der Waals potential $-C_6/R^6$. For example, in rubidium $R^{\star} \approx 10 ~ \textrm{nm}$ and $n_{\textrm{C}}= 2.5 \times 10^{17} ~ \textrm{cm}^{-3}$, about two orders of magnitude higher than typical BEC densities. For ground-state atoms colliding with ions, e.g. Rb and Ba$^{+}$, $R^{\star} \approx 300 ~ \textrm{nm}$ of the charge-induced dipole interaction $U_{\textrm{CID}} = -C_4/R^4$ is about 30 times larger. Therefore, the critical density is already reached at $n_{\textrm{C}}^{\textrm{Ba-Rb}} \approx 1 \times 10^{13} ~ \textrm{cm}^{-3}$ which is in agreement with experimental findings \cite{Haerter2012,Krukow2016,Mohammadi2021,Schmidt2020}. For Ba ions interacting with NaK molecules, $R^{\star}$ is once again drastically increased to $R^{\star} = 1.42 \times 10^{-5} ~ \textrm{m}$, such that $V^{\star}$ would be comparable to the capture volume of typical dipole traps and $n_{\textrm{C}}^{\textrm{Ba-NaK}} \approx 1 \times 10^{8} ~ \textrm{cm}^{-3}$.
	
	Following the analogy to ion-atom interactions, it would seem obvious that reaching the quantum dominated regime in the Ba$^{+}$-NaK system using even the most advanced hybrid platforms would be extremely challenging. However, it is important to note that these results strongly depend on the reduced masses and static polarizabilities of the chosen ion-molecule combination. Moreover, the Langevin model only holds in the regime of extremely dilute gases, that is $n \ll n_{\textrm{C}}$, where molecules collide with the ion one at a time, such that the intermolecular interactions are negligible. In contrast, this approximation breaks down for dense molecular gases where the dipole-dipole interactions and many-body effects are expected to play a significant role \cite{PerezRios2021}. This scenario is discussed in Section \ref{densemolecules}. Most importantly, the characteristic quantities derived so far only apply to molecules in the rovibronic ground state $\ket{J = 0, m_J= 0}$. Therefore, as a next step, we consider the influence of other rotational states on the collisional properties of ion-molecule mixtures.
	
	\subsection{Molecules in rotationally excited states}
	The effective long-range polarizability $\alpha_{0,J}^{\textrm{eff}}$ of molecules prepared in the rovibronic state $\ket{X^{1}\Sigma^{+}, v=0, J, m_J}$ contains two main contributions, $\alpha_{0,J}^{\textrm{rot}}$ arising from purely rotational transitions, and $\alpha_{0,J}^{\textrm{\textrm{el}}}$, the contribution due to vibronic transtions to excited electronic states, while the component attributed to rovibrational transitions within the electronic ground state is typically small \cite{Vexiau2017}. Since $\alpha_{0,J}^{\textrm{rot}}$ is determined by the second derivative of the potential with respect to $\mathcal{E}$, as shown in Eq.~\ref{eqn:derivatives}, both $\alpha_{0,J}^{\textrm{eff}}$ and $E^{\star}_J$ strongly depend on the choice of the rotational state. Therefore, by exploiting radiatively long-lived rotational states, all the characteristic properties that are fixed for a given combination of ions and ground-state atoms could be tuned by several orders of magnitude in polar molecules, including those with an extremely low threshold for $s$-wave collisions such as NaK and NaCs. The expected impact on the properties of ion-molecule mixtures is illustrated in Table \ref{tab:Jdependence} for two exemplary combinations, Ba$^{+}$-NaK and Ba$^{+}$-LiNa. In the following, we identify several states that may be advantageous for ion-molecule collisions in the ultracold regime, characterized by negative, negligible or low curvatures of the eigenvalues $ U_{\textrm{CD}}$ (energy levels of NaK up to $J=3$ shown in Fig. \ref{fig:eigenenergies}) with respect to $\cal{E}$ at low field which we obtain by numerically calculating $\partial^2 U_{\textrm{CD}}(\mathcal{E}) / \partial \mathcal{E}^2$. In the latter instances, $\alpha_{0,J}^{\textrm{\textrm{el}}}$ is sizable and has to be taken into account. This is in contrast to the rotational ground state where typically $\alpha_{0,0}^{\textrm{\textrm{el}}} \ll \alpha_{0,0}^{\textrm{rot}}$.

	\begin{table}[b]
		\caption{\label{tabtwo} Effective static polarizabilities $\alpha_{0,J}^{\textrm{eff}}$ and characteristic energies $E^{\star}_J$ for collisions with Ba$^{+}$ for selected rotational states $\ket{J,m_J}$. The low-field seeking state $\ket{J=1, m_J=0}$ yields imaginary $E^{\star}_J$.} 
		\begin{ruledtabular}
			\begin{tabular} {rldd} 
				& rot. state &   \multicolumn{1}{c}{$\alpha_{0,J}^{\textrm{eff}}$} &   \multicolumn{1}{c}{$E^{\star}_{J}/k_{\textrm{B}}$} \cr 
				&  \multicolumn{1}{c}{$\ket{J, m_J}$} &    \multicolumn{1}{c}{$(10^{-37}~ \textrm{C m}^2 \textrm{V}^{-1})$} &    \multicolumn{1}{c}{(nK)} \cr
				\colrule
				\multirow{5}{*}{NaK} & $\ket{0, 0}$ & 152 & 0.0281 \\
				& $\ket{1, 0}$  & -91.1 &  \\
				& $\ket{1, \pm1}$  & 45.6 & 0.0939 \\
				& $\ket{2, \pm2}$ & 21.7 & 0.197 \\
				& $\ket{3, \pm2}$  & 0.0578 & 74.1 \\
				\colrule
				\multirow{6}{*}{LiNa} & $\ket{0, 0}$ & 1.62 & 7.86 \\
				& $\ket{1, 0}$  & -0.935 &  \\
				& $\ket{1, \pm1}$  & 0.487 & 26.512 \\
				& $\ket{2, \pm2}$ & 0.232 & 55.676 \\
				& $\ket{3, \pm2}$  & 0.0385 & 335 \\
				& $\ket{5, \pm1}$  & 0.00109 & 11900 \\
				
			\end{tabular}
		\end{ruledtabular}
		
		\label{tab:Jdependence}
	\end{table}
	One interesting example from this class is %
	the first rotationally excited state $\ket{X^{1}\Sigma^{+}, v=0, J = 1, m_J=0}$ which features a reversed curvature as shown in Fig.~\ref{fig:polarization}(b) for NaK. This means that the polarization potential is rendered repulsive with a barrier height of $U_{0}^{(J = 1, m_J=0)} /k_{\textrm{B}} = 87.4 ~ \textrm{mK}$ reached around \linebreak $\mathcal{E} = 10 ~ \textrm{kV cm}^{-1} \approx  5~ \mathcal{E}^{\star}$ in NaK as depicted in Fig.~\ref{fig:eigenenergies}.  Therefore, for molecules prepared in this state and at typically achievable temperatures on the order of $T_{\textrm{mol}} \approx 100~ \textrm{nK}$ much lower than $U_{0}^{(J = 1, m_J=0)} / k_{\textrm{B}}$, we expect that short-range ion-molecule collisions, and ion-molecule-molecule TBR, will be strongly suppressed, while many desired effects such as sympathetic cooling are still expected to be efficient. We note, that this expectation holds even in presence of ion micromotion and despite the prediction that $E^{\star}$ in the rovibronic ground state is extremely low compared to the thermal energy of typical molecular quantum ensembles. In ion-atom systems, conceptually similar ideas for designing effectively repulsive potentials by exploiting Rydberg dressing induced by optical fields have been proposed \cite{Secker2017}. In this case, a potential barrier arises from the competition of an engineered repulsive contribution and the attractive static charge-induced dipole potential, both scaling as $R^{-4}$. In contrast, for molecules interacting with ions, the long-range potential itself is repulsive such that the sought-after shielding can be facilitated by exploiting their rich energy level structure, that is, without additional dressing.
	
	As a second example we consider LiNa prepared in the state $\ket{J = 5, m_J= \pm 1}$. We find that the polarizability contribution calculated from its curvature is opposite in sign and comparable in magnitude to that stemming from coupling to electronically excited states \cite{Vexiau2017}. The effective polarizability is then about $10^{3}$ times smaller compared to $\ket{J = 0, m_J= 0}$, with the corresponding $E_J^{\star} / k_{\textrm{B}} \approx 10 ~ \mu\text{K}$ being comparable to the most favorable ion-atom mixtures in rf-hybrid traps. For NaK, we find no states with such properties for rotational manifolds up to $J=7$. This suggests that entering the quantum regime of ion-molecule collisions may be feasible, with  methodology similar to that established in ion-atom systems, for suitable combinations with especially low potential curvatures such as Ba$^{+}$-LiNa.
	
	Another instructive case is the state $\ket{J = 3, m_J= \pm 2}$. We calculate the derivatives of its potential with respect to $\cal{E}$ for different molecules and find that the curvature is negligible compared to that of $\ket{J = 0, m_J= 0}$. In the exemplary combinations of Ba$^{+}$ with NaK and LiNa this leads to a drastic increase of the energy threshold to $E_J^{\star}(\text{NaK}) /k_{\textrm{B}} = 74 ~ \text{nK}$ and $E_J^{\star}(\text{LiNa}) /k_{\textrm{B}} = 335 ~ \text{nK}$, respectively. We find similar results in CaH where we estimate $E_J^{\star}(\text{CaH}) /k_{\textrm{B}} \approx 900 ~ \text{nK}$ using $\alpha_0^{\textrm{rot}} \approx d_0^2/(3 B_0)$ \cite{Lepers2013} and $\alpha_0^{\textrm{\textrm{el}}}$ derived from an Unsöld approximation \cite{Julienne2011} with a theoretically predicted long-range van der Waals coefficient for the $\ket{X^{2} \Sigma^{+}}$ ground state of $C_6 = 119 ~\text{a. u.}$ \cite{Weck2003}. %
	When considering lighter or heavier ions, e.g. $^{174}$Yb$^+$, $^{88}$Sr$^+$, $^{40}$Ca$^+$, $^{24}$Mg$^+$ or $^{9}$Be$^+$ combined with NaK, LiNa and CaH, Eq.~\ref{eqn:Rstar} and Eq.~\ref{eqn:Estar} predict a scaling as $R^{\star} \propto \sqrt{\alpha_{0,J}^{\textrm{eff}} \mu} $ and $E^{\star} \propto 1/(\alpha_{0,J}^{\textrm{eff}} \mu^{2}) $, with the resulting characteristic energies ranging from $E_J^{\star} /k_{\textrm{B}} = 65 ~ \text{nK}$ to approximately $16 ~ \mu\text{K}$ as summarized in Table \ref{tabIonMolComb}. 

In these examples, the energy scale for $s$-wave collisions is comparable to typical ion-atom mixtures \cite{Cetina2012,Tomza2019}. There the currently best experiments in rf-based hybrid traps are capable of mitigating micromotion-induced heating to a point where the collision energies reach $E_{\textrm{col}} / k_{\textrm{B}} \approx 10 ~\mu\text{K}$ which approaches the $s$-wave limit for the most favorable combinations. On the other hand, optical hybrid traps in principle allow for sympathetic cooling of any suitable ionic species to the temperature of the neutral ensemble. Assuming thermalization between a Ba$^{+}$ ion and molecular quantum gases with now reachable temperatures on the order of $10 ~ \text{nK}$ \cite{Schindewolf2022,Bigagli2024}, this means that, for example, $E_{\textrm{col}} \approx E^{\star}_J /5$ for NaK and $E_{\textrm{col}} \approx E^{\star}_J /22$ for LiNa could be achieved, reaching into the quantum dominated regime. For this choice of the rotational state, ion-molecule experiments with collision energies on the order of $E_J^{\star}$ would likely require the capability to optically trap the ions and the molecules at the same time. On the other hand, studies of cold and ultracold chemical reactions and their additional control obtained by manipulating rotational degrees of freedom, which are the focus of Section \ref{reactive collisions}, or other experiments involving $E_{\textrm{col}}/k_{\textrm{B}}$ on the order of mK or above could be carried out using rf-based hybrid traps. In the following section, we discuss the methodological requirements and feasibility of such hybrid ion-molecule platforms with respect to the currently available approaches and techniques.
\begin{table}
	\caption{\label{tabIonMolComb} Characteristic energy scale $E_J^{\star}/k_{\textrm{B}} ~ \textrm{(nK)}$ for different combinations of ions and molecules prepared in the rotational state $\ket{J=3, m_J= \pm 2}$.} 
	\begin{ruledtabular}
		\begin{tabular} {rccc}
			&\text{NaK} &  \text{LiNa} &  \text{CaH} \cr
			\colrule
			\multirow{1}{*}{$^{174}$Yb$^{+}$} & 65 & 311 & 803 \\
			\multirow{1}{*}{$^{138}$Ba$^{+}$} & 74 & 335 & 885  \\
			\multirow{1}{*}{$^{88}$Sr$^{+}$} & 102 & 407 & 1130 \\
			\multirow{1}{*}{$^{40}$Ca$^{+}$} & 229 & 692 & 2160 \\
			\multirow{1}{*}{$^{24}$Mg$^{+}$} & 453 & 1145 & 3860  \\
			\multirow{1}{*}{$^{9}$Be$^{+}$} & 2190 & 4246 & 16200
		\end{tabular}
	\end{ruledtabular}
\end{table}
	\section{Experimental toolbox for hybrid trapping of ions and neutral molecules} \label{toolbox}
	\begin{figure*}[!htb]
		\centering
		\includegraphics[width=0.85\linewidth]{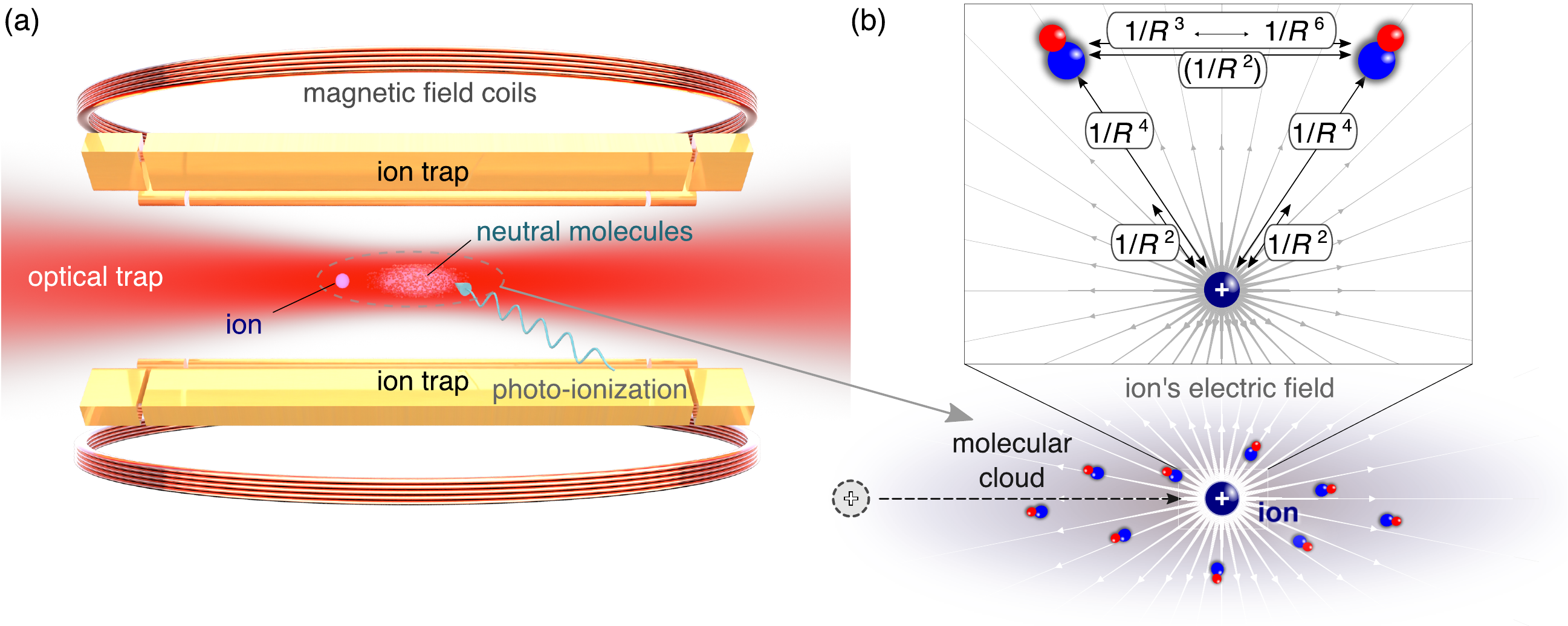}
		\caption[MPHT]{\textbf{(a)} Schematic of an ion-molecule hybrid trap combining a magnetic trap, optical dipole traps, and an integrated trap for ions for investigating and controlling the interactions between single ions and neutral molecular quantum gases. \textbf{(b)} An ion immersed into the neutral cloud introduces a source of static electric fields, aligning the permanent molecular electric dipoles, depending on the distance $R$. This provides access to a new system with a rich hierarchy of competing tunable interactions spanning 1/$R ^{6} $ (induced dipole-dipole, attractive), 1/$R ^{4} $ (charge-induced dipole, attractive or repulsive: depending on the quantum numbers $ J, m_J$), 1/$R ^{3} $ (dipole-dipole, repulsive for side-by-side or attractive for head-to-tail collisions), 1/$R ^{2} $ (charge-dipole, attractive or repulsive: determined by the electric field and $ J, m_J$). The transition from the 1/$R ^{4} $ to the 1/$R ^{2} $ potential occurs above the characteristic electric field $\mathcal{E}^{\star}$.}
		\label{fig:mphtandimi}
	\end{figure*}
	Modern experiments capable of producing ultracold ground-state bialkali molecules follow a route building upon methodology developed for the preparation of quantum degenerate atomic gases, extended to facilitate a combination of two species in the same apparatus. This typically involves Zeeman slowing of thermal beams and subsequent laser cooling, magneto-optical, magnetic, and optical trapping followed by evaporative cooling (close) to quantum degeneracy. The final steps include the association of loosely-bound molecules by exploiting magnetically tunable Fano-Feshbach resonances, and finally, the controlled transfer to the rovibronic ground state via stimulated Raman adiabatic passage (STIRAP), a coherent two-photon process. When considering the part required for the creation of an ultracold ensemble, a typical molecule apparatus incorporates additional layers of complexity compared to an experimental setup designed for trapping and manipulation of a single atomic species. Nonetheless, several experiments have succeeded in preparing and controlling ensembles of ground-state molecules, with the achievement of quantum-degenerate molecular gases marking a milestone in the development of the field \cite{DeMarco2019,Schindewolf2022,Bigagli2024}. Despite the additional complexity, such experiments can be realized in a laboratory environment that is comparable to that of single-species machines.
	
	In parallel to these developments in the field of ultracold molecular gases, first experiments for studying ultracold ion-atom collisions have been realized \cite{Smith2005,Grier2009}. Their aim was to combine vastly different methodologies for preparing and controlling atoms and ions at ultralow temperatures in a common two-species apparatus. Such hybrid setups using radiofrequency (rf)-driven Paul traps to confine ions and (magneto-) optical traps for preparing and controlling ultracold atomic ensembles have proven to be a versatile platform and have been successfully demonstrated for many ion-atom combinations \cite{Smith2005,Grier2009,Schmid2010,Zipkes2010,Ravi2012,Hall2013,Rellergert2011,Mohammadi2021,Haze2018,Meir2016,Deiglmayr2012,Rahaman2024,Weckesser2021,Feldker2020,Li2020}. More recently, it was shown that this concept can be extended to combinations of several species of atomic ions and multi-species atomic magneto-optical traps which can be used to study ion-atom collisions at low temperatures and in principle support the association of bialkali molecules \cite{Rahaman2024}. Among the many important developments in this field, we point out two specific examples that are most relevant in view of the proposed extension of the study of ion-neutral interactions to ensembles of molecular quantum gases, that is, highly optimized rf-based hybrid traps and all-optical ion-atom traps. In the former case, for favorable ion-atom combinations featuring low reduced masses and long-range induction coefficients, optimized Paul traps in conjunction with advanced schemes for the detection and compensation of stray electric fields have allowed to reduce the collision energy to a range where non-classical effects become visible \cite{Feldker2020,Weckesser2021}. In the latter case, similarly optimized setups were also combined with bichromatic optical traps capable of simultaneously confining ions and atoms \cite{Schmidt2020,Karpa2021}. In particular, using optical traps provides the advantage that micromotion-induced heating can be completely eliminated and that this method is applicable to generic combinations of species provided that an effective optical trapping potential can be engineered \cite{Karpa2019}. In principle, this is the case for any combination of ions and neutral atoms or molecules possessing a non-zero dipole moment.
	
	Therefore, we expect that a similar approach to overlapping ions and neutral species, that is, either using rf-based or all-optical hybrid traps, can also be applied to neutral molecules. In addition to ensembles of bialkali molecules such as those given in Table \ref{tab:moleculeoverview}, this should also be the case e.g. for single molecules trapped in optical tweezers \cite{Zhang2022a,Gregory2024,Vilas2024,Holland2023}, other classes of polar molecules such as directly laser-coolable CaH \cite{VazquezCarson2022}, CaF \cite{Holland2023}, SrF \cite{Langin2021}, BaF \cite{Zeng2024}, and YO \cite{Burau2024}, or polyatomic molecules such as CaOH \cite{Vilas2024} and CaOCH$_{3}$ \cite{Mitra2020}. For clarity, we will focus our discussion on the exemplary ensembles of ground-state NaK and Ba$^{+}$ ions for the following reasons. First, molecular quantum (-degenerate) gases with a high phase-space density have been recently achieved for NaK \cite{Voges2020,Schindewolf2022}, and second, both the ions and the molecules can be routinely confined and prepared in standard optical dipole traps \cite{Huber2014,Lambrecht2017, Schmidt2018,Hoenig2024,Voges2020,Schindewolf2022}. Furthermore, as we will discuss in the remainder of this section, NaK molecules prepared in the rovibronic $^{1}\Sigma^{+}$ ground state have very similar properties to common alkali species regarding mass and dynamic dipole polarizability $\alpha_d(\omega)$, with $\omega$ denoting the frequency of the interacting optical field. This allows for detailed comparisons of the expected behavior of the proposed ion-molecule system with Ba$^{+}$-Rb mixtures which have been extensively studied both in presence of micromotion-induced heating encountered in rf-hybrid \cite{Haerter2012,Krukow2016,Mohammadi2021} and in rf-free optical traps where micromotion is eliminated \cite{Schmidt2020}. Finally, near-infrared (NIR) far-off resonant optical traps operated at a wavelength of 1064 nm, have been successfully used both to trap NaK and to demonstrate a generic approach to studying ion-atom collisions, first realized for Ba$^{+}$-Rb mixtures \cite{Schmidt2020}. To this end, bichromatic optical potentials are created by overlapping dipole traps of different wavelengths.
	
\begin{figure}[!htb]
		\centering
		\includegraphics[width=0.99\linewidth]{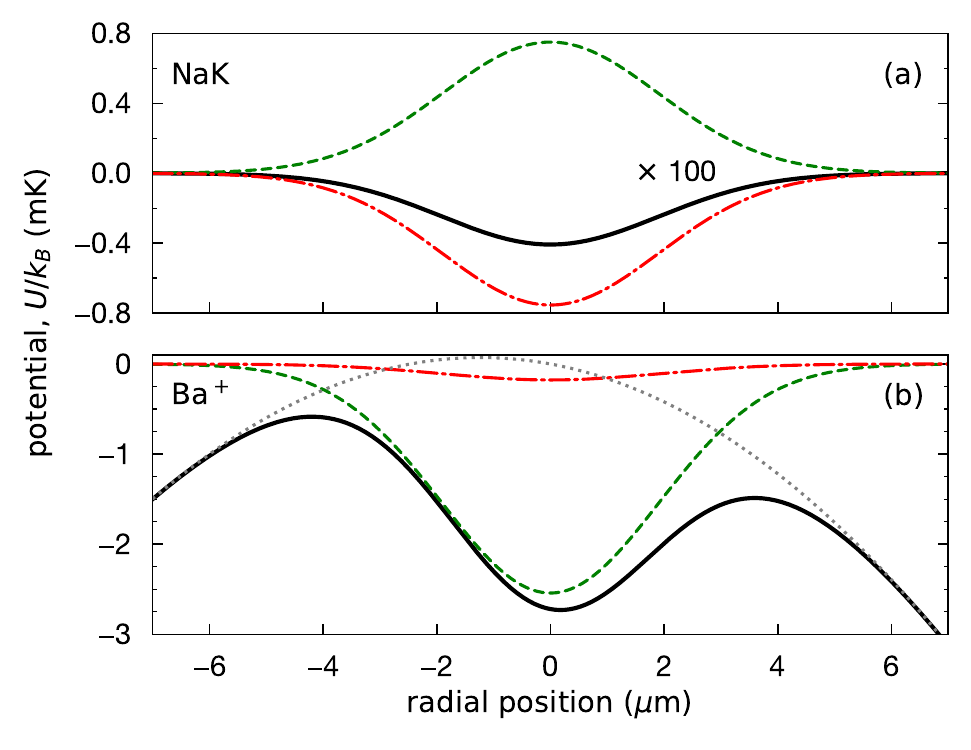}
		\caption[bichroODT]{\textbf{(a)} Effective bichromatic radial (orthogonal to the ODT's wavevector) potential (black solid line, enlarged by a factor of 100 for visibility) for NaK (in the rovibronic ground state $ \ket{\textit{X} ^{1}\Sigma^{+}, v = 0, J = 0} $) exposed to a composite optical dipole trap superimposing ODTs at 1064 nm (NIR, red dashed-dotted line) and at 532 nm (VIS, green dashed line). \textbf{(b)} The corresponding potentials for Ba$^{+}$ (in the electronic ground state $6^{2}S_{1/2}$). The asymmetry and negative curvature of the effective potential at positions $\gtrapprox 4 ~\mu\text{m}$ around the trap center are caused by a residual stray electric field and the electrostatic potential of the Paul trap (gray dotted line: radial component), respectively. The latter is kept active during the optical trapping phase causing radial defocussing. The Paul trap and ODT parameters are adopted from experiments with Ba$^{+}$-Rb mixtures \cite{Schmidt2020}, accounting for the properties of NaK at a temperature of $T^{\textrm{NaK}} = 400 ~ \textrm{nK}$.}
		\label{fig:bichroODT}
	\end{figure}

	For a direct comparison with previous experiments we use the parameters employed in \cite{Schmidt2020}, that is, optical dipole traps with wavelengths of 532 nm (VIS ODT) and 1064 nm (NIR ODT), optical powers $P_{\textrm{VIS}} = 0.47 ~ \textrm{W}$ and $P_{\textrm{NIR}} = 0.13 ~ \textrm{W}$ (adjusted for typical temperatures of NaK quantum gases), matched $1/e^2 $ waist radii $w_0^{\textrm{VIS}} = w_0^{\textrm{NIR}} = 3.8 ~ \mu \textrm{m}$, a residual stray electric field of $\mathcal{E}_{str} = 10 ~ \textrm{mV} ~\textrm{m}^{-1}$, and the axial secular frequency of the linear Paul trap of $\omega_{x} / (2 \pi) = 12 ~\textrm{kHz}$, respectively. We then calculate the bichromatic potential suitable for overlapping a barium ion laser-cooled close to the Doppler temperature $T_{\textrm{D}}^{\textrm{Ba}} = 365 ~ \mu\textrm{K}$ with an ensemble of ground-state NaK molecules prepared at $T^{\textrm{NaK}} = 400 ~ \textrm{nK}$. The NaK molecules experience a repulsive potential from the VIS ODT due to its blue detuning with respect to the energetically lowest dipole-allowed optical transition $X ^{1}\Sigma^{+} \leftrightarrow A^{1}\Sigma^{+}$. To maintain the initial trap depth $U_0^{\textrm{NaK}} /(k_{\textrm{B}}) \approx 4 ~ \mu\textrm{K}$, the intensity of the NIR ODT has to be increased compared to its initial value in order to achieve the desired differential light shift. Despite the substantially more complex energy level structure of molecules in comparison with alkali atoms, the resulting potentials calculated in the far-off-resonant limit \cite{Grimm2000} and using the dynamic dipole polarizability of NaK \cite{Vexiau2017} that are shown in Fig.~\ref{fig:bichroODT} are quantitatively very similar to those obtained for the previously demonstrated case of Ba$^{+}$-Rb in terms of trap depths and ODT parameters. This suggests that the protocols for overlapping ions and atoms demonstrated in \cite{Schmidt2020} as well as proposed methods for further improving the trapping performance and realizing more complex trapping configurations \cite{Schmidt2018,Karpa2019,Hoenig2024} are readily applicable to molecules. The expected compatibility to such an extent may seem surprising at first glance because, in general, neutral atoms and polar molecules may have vastly different polarizabilities. This is indeed the case for the static polarizability $\alpha_0$ which is a pivotal point of the proposed experiments and is discussed in detail in Section \ref{molecules_approaching_ions}. However, in the optical and near-infrared frequency domains where the detunings with respect to rotational transitions are several orders of magnitude larger than those for optical transitions, all $\alpha_d(\omega)$ for the bialkali molecules in the rovibronic ground state as listed in Table \ref{tab:moleculeoverview} vary by less than a factor of 2 and are of the same order of magnitude as those of common alkali atoms \cite{Vexiau2017}. It is this robustness of the dynamic polarizability that facilitates the creation of molecules from individually prepared atomic species which are typically all confined in the same far-detuned optical dipole traps. At the same time, even within the same electronic and vibrational ground state manifold, i.e. $ \ket{\textit{X} ^{1}\Sigma^{+}, v = 0} $, $\alpha_d(\omega)$ exhibits a substantial dependence on the rotational quantum numbers $J$ and $m_J$ \cite{Bause2020}.

	Exploiting the similarities between molecules and alkali atoms interacting with far-detuned optical fields, a prototypical experimental setup could be realized as shown in Fig.~\ref{fig:mphtandimi}. Its central features include (i) a Paul trap optimized for ion-neutral experiments, that is, optical access through high-numerical-aperture objectives beneficial for stray electric field compensation and focusing of the dipole trap beam(s) and the correspondingly designed ion-electrode spacing $(\approx 1 ~ \textrm{mm})$, (ii) magnetic field coils required for the magneto-association of Feshbach molecules, as well as (iii) a bichromatic optical dipole trap. Similar Paul traps have been demonstrated and feature suppression of stray electric fields down to $\approx 10^{-3} ~ \textrm{V m}^{-1}$ \cite{Schmidt2020,Weckesser2021,Weckesser21b,Hoenig2024}, very low anomalous heating rates \cite{Lambrecht2017}, and typical trap depths on the order of $k_{\textrm{B}} \times 10^{4}~\textrm{K}$, which is sufficiently deep to recapture ionic products.
	
	The considerations above suggest that the proposed hybrid platform is suitable for investigations of ultracold ion-molecule interactions, such as precise studies of controlled quantum chemistry and ultracold collision experiments with dense molecular gases which are discussed in the following Sections \ref{reactive collisions} and \ref{densemolecules}, respectively.
	
	\section{Reactive collisions and formation of molecular ions} \label{reactive collisions}
	
	One of the most striking features of ion-molecule collisions setting them apart from those between atoms and ions is the much larger static polarizability of heteronuclear molecules, which translates to a drastically lower threshold energy $E^{\star}$. This means that even at the lowest temperatures on the order of $10 ~ \textrm{nK}$ that have been observed in molecular systems to date \cite{Valtolina2020,Schindewolf2022,Bigagli2024}, the collisions of all ground-state molecules given in Table \ref{tab:moleculeoverview} (except for LiNa) with ions would take place in the classical regime. In this limit, the cross section of capture processes is well described by the Langevin model, which predicts $\sigma_{\textrm{L}} =2 \pi \sqrt{C_4/E_\textrm{col}}$. The related rate constant for Langevin collisions is independent of $E_\textrm{col}$: $K_{\textrm{L}} = \sigma_{\textrm{L}}  \hbar k /\mu = \pi \sqrt{8 C_4 / \mu}$. For a direct comparison we consider the interaction of a $^{138}\textrm{Ba}^{+}$ ion with $^{23}\textrm{Na}^{39}\textrm{K}$ molecules. With the strongly enlarged polarizability of NaK, we expect that the rate for reactive collisions will be enhanced by a factor of $K_{\textrm{L}}^{\textrm{Ba-NaK}} / K_{\textrm{L}} ^{\textrm{Ba-Rb}} \approx 60$, which is typical for most bialkali molecules.

	\begin{figure}[b]
		\centering{%
			\includegraphics[width = 1.0 \linewidth]{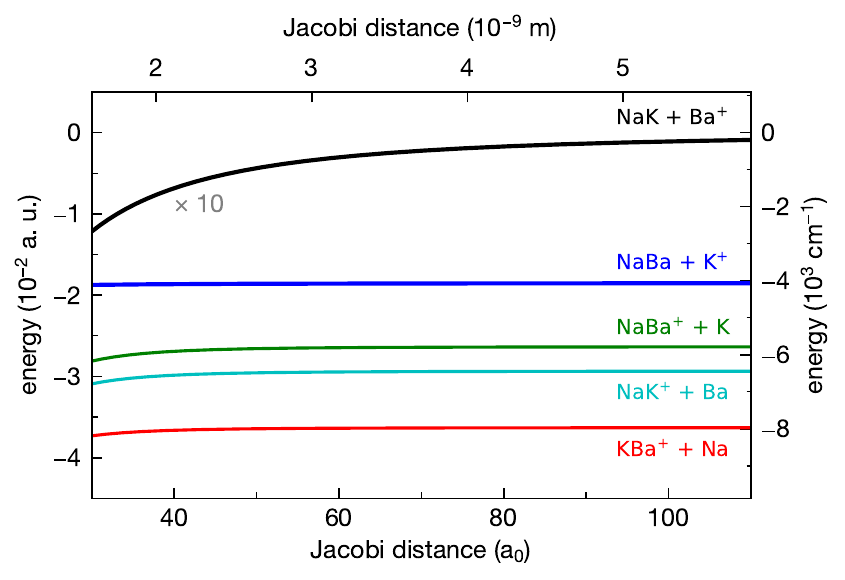}%
		}
		\caption{
			Energies of possible combinations of ionic (and neutral) diatomics (energy taken at the minimum of their ground state PEC) and ground-state neutral atoms (and ions) in a mixture of Ba$ ^{+} $ and NaK (entrance scattering channel), as a function of the ion-molecule (Jacobi) distance $R$ (courtesy of Romain Vexiau). Reactive and charge exchange channels are energetically well separated from the prepared ensemble of Ba$ ^{+} $ and NaK at $ R > 60 ~a_0 $ ($a_0$: Bohr radius).  The long-range variation of the PECs strongly depends on the combination. All energies have been scaled up by a factor of 10 around their asymptotic values for visibility.
		}
		\label{fig:NaKBaReactive}
	\end{figure}
	\begin{table}
		\caption{\label{tabthree} Dependence of the Langevin cross section $\sigma_{\textrm{L}}^{\textrm{NaK}}$ (for a Ba$^{+}$ ion cooled to the Doppler limit of $370~ \mu\text{K}$ and NaK at $400~ \text{nK}$) and reactive rates $K_{\textrm{L}}^{\textrm{Ba-NaK}} $ on selected rotational vibronic ground states $\ket{J,m_J}$. $K_{\textrm{L}}^{\textrm{Ba-NaK}} $ is normalized to the case of Ba$^{+}$-Rb.} 
		\begin{ruledtabular}
			\begin{tabular} {ldd}
				rot. state &   \multicolumn{1}{c}{$\sigma_{\textrm{L}}^{\textrm{NaK}} ~ (10^{-24} ~\text{m}^{2})$} &   \multicolumn{1}{c}{$K_{\textrm{L}}^{\textrm{Ba-NaK}} / K_{\textrm{L}} ^{\textrm{Ba-Rb}}$ }\cr 
				\colrule
				$\ket{0, 0}$ & 1.55 & 60 \\
				$\ket{1, 0}$ & 0 & 0 \\
				$\ket{1, \pm 1}$ & 0.85 & 32.9 \\
				$\ket{2, \pm 2}$ & 0.587  & 22.7 \\
				$\ket{3, \pm 2}$ & 0.0303 & 1.17 \\
				
			\end{tabular}
		\end{ruledtabular}
		\label{tab:reactivities}
	\end{table}

	In ion-atom systems the reactive capture cross sections show a very strong dependence on the electronic state of both the ion and the atomic gas. For example, in collisions between Yb$^{+}$ and Li, the loss rates of ions in the states $ ^{2}S_{1/2}, ~ ^{2}P_{1/2}, ~  ^{2}D_{3/2}$, and $^{2}F_{7/2}$ vary by several orders of magnitude, with the lowest reactivity observed in $^{2}S_{1/2}$ \cite{Joger2017}. Similarly, $K_{\textrm{L}}$ of $^{2}S_{1/2}$ Yb ions interacting with Li atoms in the Rydberg state $24 S$ is about $10^3$ times larger than for ground-state atoms \cite{Ewald2019}. In a similar manner, the polarizability of molecules can be manipulated by preparing a specific state $\ket{J, m_J}$, as illustrated in Table \ref{tab:reactivities}. One difference is that these changes are expected without excitation from the electronic ground state or even the vibrational state within the same manifold, that is, they can be brought about with microwave fields. Another distinctive feature is that the rotational states are long-lived compared to electronically excited states \cite{Ruttley2025}. Finally, the sign of the polarizability itself can be changed by selecting states with specific $m_J$. This means that short-range collisions which are strongly enhanced in the ground state can be suppressed such that using the degrees of freedom offered by the molecules the reactive rates can be tuned from zero to extremely high values comparable to or exceeding those achievable in mixtures of ions and Rydberg atoms. An interesting application of this sensitivity would be to use chemical reactions with the ion as a probe of collisional processes within molecular ensembles %
	or between molecules and atoms \cite{Yang2019, Son2020, Gregory2021, Nichols2022, Voges2022}. Exploiting the strong dependence of the reaction rates on the initial state of the molecules could provide additional insight into elusive effects such as the formation of collisional complexes. One possible detection scheme would be to prepare NaK in $ \ket{ J = 1, m_J = 0} $ (with negligible $\sigma_{\textrm{L}}$ for Ba$^{+}$). The intermediate tetramer complex is very likely to be in a different rotational state where an enhanced reactivity is expected. Thus, the detection of ion loss would signal the presence of such complexes and may be used to extract information about their creation rates and lifetimes.
	\begin{figure}[!htb]
		\includegraphics[width=1.0\linewidth]{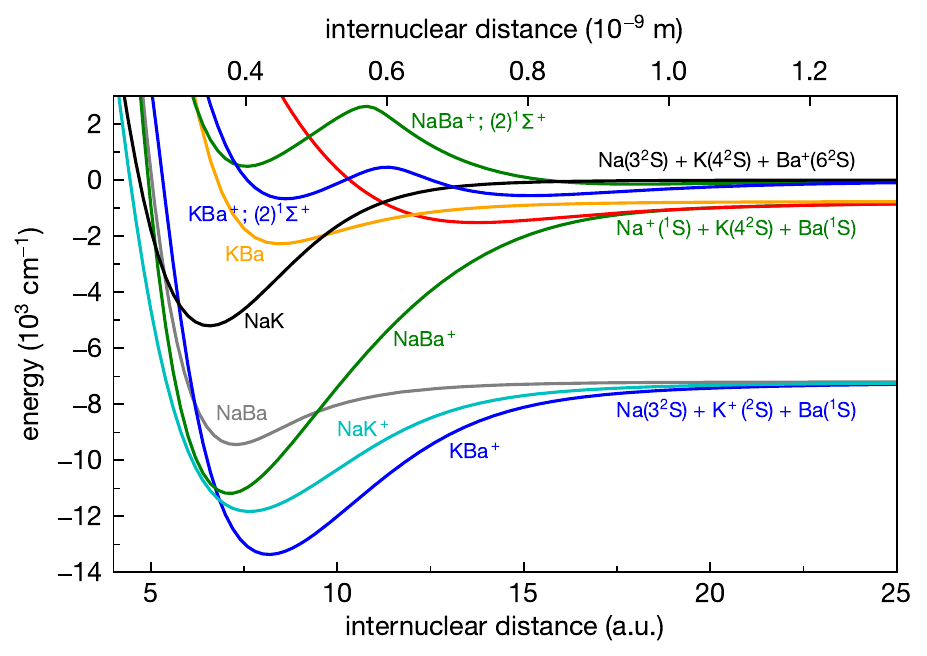}
		\caption{
			Calculated energies of possible ground state ($X^1\Sigma_g^+$) diatomic molecules composed of Na, K neutral atoms and a Ba$^{+}$ ion referenced
			to the ground-state energy of these three particles. The calculations are performed with the
			methodology of \cite{Aymar2005,Guerout2010,Vexiau2017} (courtesy of Romain Vexiau). Note that for completeness we display the second excited state of NaBa$^{+}$ and KBa$^{+}$ of the same symmetry as they dissociate towards the same limit. The formation of the dimers NaBa$^{+}$, NaK$^{+}$, and KBa$^{+}$ that are trappable in the same Paul trap as the initially prepared Ba$^{+}$ ions is energetically possible. The branching ratios into the different channels as well as the rates of elastic and inelastic collisions strongly depend on the spin and hyperfine states of the constituents and provide a system with rich chemical and collisional properties that can be explored in the proposed setup. The presence of an ion further opens the possibility to control the collisional properties of polar NaK molecules, or NaKBa$ ^{+} $ trimer formation \cite{Pandey2024}, as well as the impact of the ion on the potential creation of NaK-NaK tetramer complexes or, assisted by TBR, ionic pentamers like Na$ _{2} $K$ _{2} $Ba$ ^{+} $.
		}
		\label{fig:NaKBaGround}
	\end{figure}

	In order to obtain systems with extremely high chemical reactivity that can be controlled on the quantum level, in addition to a suitably high polarizability, we also require energetically allowed target channels \cite{Smiakowski2020}. To investigate if this is the case for the combination of Ba$^{+}$ and NaK, we calculate the potential energy surfaces for all combinations of neutral and molecular reactants. As shown in Fig. \ref{fig:NaKBaReactive} and \ref{fig:NaKBaGround}, we indeed find several open reactive channels. We expect the production of ionic dimers NaK$^{+}$, NaBa$^{+}$, and KBa$^{+}$ as well as the atomic ions K$^{+}$, Na$^{+}$, and Ba$^{+}$ and all neutral constituents. This  indicates that the system is highly suitable for studies of quantum chemistry at low and ultralow collision energies. However, the predicted energy release of the exoergic reactions is on the order of $k_{\textrm{B}} \times 10^{4} ~ \textrm{K}$. In neutral molecule systems reactions on this energy scale are predominantly signaled by loss. Despite this complication, the rapid development in the field has allowed to obtain much more information about the collisional processes using kinetic energy spectrometry based on photo-ionization in conjunction with velocity-map imaging \cite{Hu2019}.
	
	In the present case, a hybrid ion-molecule apparatus would be equipped with an integrated Paul trap, such that, in principle, the expected ionic reaction products may be recaptured provided that the trap is suitable for this task. Typical Paul traps, including those employed in hybrid experiments, reach well depths on the order of $k_{\textrm{B}} \times 10^{4} ~ \textrm{K}$ \cite{Leibfried2003} which is comparable to the kinetic energy of the expected products. Moreover, the stability region of Paul traps for $\textrm{Ba}^{+}$ can be engineered to support the charge-to-mass ratios of the expected product ions which are similar to $\textrm{Rb}^{+}$ and $\textrm{Rb}_2^{+}$, species that have been sympathetically cooled by cotrapped $\textrm{Ba}^{+}$  \cite{Schmidt2020,Schmidt2020b}. Exploiting these properties, barrierless ion-neutral reactions may provide an elegant way to synthesize and study polyatomic molecular ions in hybrid traps \cite{Puri2017}. In this regard, similar systems may be realized either in successive ion-molecule collision experiments or by irradiating the ensembles with an ultraviolet photoionization beam, in analogy to recent studies of collisions between heteronuclear dimers \cite{Hu2019}. Furthermore, it was shown that cooling of ions in a neutral buffer gas is extremely efficient in the classical collision regime and at high energies, even if the ions are not laser-coolable \cite{Ravi2012,Mohammadi2021}. In our case, ultracold and dense ensembles of three different neutral species, that is Na, K, and NaK can be used for this purpose. Therefore, it is likely that the molecular ions produced in ion-molecule reactions may be recaptured allowing for further manipulation and analysis. This includes sympathetic cooling \cite{Meir2018}, high-resolution mass spectrometry \cite{Schmidt2020b}, and (single-shot) Doppler cooling thermometry (SS)DCT \cite{Meir2017}. Ultimately, the produced molecules may also be investigated in high-precision measurements based on adapted quantum logic spectroscopy (QLS) methods \cite{Wolf2016,Sinhal2020}.
	\section{Ions and high phase-space density polar molecules} \label{densemolecules}
	\subsection{Role of dipole-dipole interactions} \label{dipoleinteractions}
	Recent experiments capable of reaching ultralow ion-atom collision energies have also enabled first studies of interactions between atomic ions and loosely bound heteronuclear molecules \cite{Hirzler2022}. Taking this concept further to a configuration where ions are immersed into a bath of polar molecules may allow exploring a new regime of interactions due to the important role that the alignment of the molecules is expected to play in the dynamical evolution of such mixtures \cite{PerezRios2021}.
	
	In the following, we discuss the prospects for such experiments, based on the known properties of collisions between ions and neutral atoms and take into account the influence of the ion's electric field on the expected effective potentials. 
	Here we include the impact of the particle density and the internal energy structure of the molecules in their rovibronic ground state
	to derive the expected changes of the interaction potential when approaching the immersed ion.
	
	With the development of methods for the shielding of undesired inelastic collisions, polar molecule experiments are now capable of producing quantum degenerate ensembles \cite{DeMarco2019,Schindewolf2022,Bigagli2024} with high phase-space and absolute particle densities reaching the order of magnitude of 1 and $10^{13} ~\textrm{cm}^{-3}$, respectively. Under such conditions, the interparticle distance is typically $\approx 100 ~ \textrm{nm}$, which is below the range of charge-induced dipole interactions in the ion-molecule system. As a consequence, the attractive potential will be experienced not by a single molecule but by an ensemble contained within the active volume $V^{\star}$. We therefore expect that in this situation several molecules will approach the ion at the same time. As an example, for NaK ensembles interacting with a Ba$^{+}$ ion $V^{\star}\textrm{(NaK)} = 1.33 \times 10^{-8} ~ \textrm{cm}^{3}$ such that this scenario occurs at comparatively low densities, e.g. at $n \approx 10^{10} ~\textrm{cm}^{-3}$ on the order of 100 molecules would be addressed simultaneously. 
	
	In stark contrast to atoms, whose collisions with ions are dominated by charge-induced dipole interaction $U_{\textrm{CID}}=-C_4/R^4$, molecules in proximity of the ion become strongly aligned along the electric field axis. Their energy shifts are then dominated by the linear Stark effect. In the radially symmetric electric field, this gives rise to additional contributions to the effective potential due to the repulsive dipole-dipole interaction between pairs of molecules labeled as $i, j$ described by the following Hamiltonian:
	\begin{equation}
		\hat{H}_{\textrm{DD}} = \sum_{\textrm{i} \neq \textrm{j}} \frac{\hat{d}_{\textrm{i}} \hat{d}_{\textrm{j}} \left( 1 - 3 \cos^2(\varphi)  \right) } {| \vec{R}_{\textrm{ij}} |^3},
		\label{eqn:DDinteraction}
	\end{equation}
	where $\varphi$ is the angle between the dipole moments $\hat{d}_{\textrm{i,j}}$ and $\vec{R}_{\textrm{ij}}$ the vector connecting the centers of mass of the two molecules. At distances where $\mathcal{E}$ is on the order of magnitude of $\mathcal{E}^{\star} $, the effective dipole moment is a sizable fraction of $d_0$ and the dominant ion-molecule interaction transitions from the attractive (in the rovibronic ground state) $U_{\textrm{CID}}$ in the long range to $U_{\textrm{CD}} \approx -d_0 \mathcal{E}(R) \propto -d_0/R^2$ in vicinity of the ion where the dipole moments are almost fully aligned with respect to $\vec{\mathcal{E}}$. For example, in rovibronic ground-state NaK molecules the effective dipole moment amounts to $0.67~ d_0$ at $\mathcal{E} = 5 ~ \mathcal{E}^{\star} $ as shown in Fig.~\ref{fig:polarization}.
	
\begin{figure}[b]
		\begin{center}
			\includegraphics[width= 0.5 \textwidth]{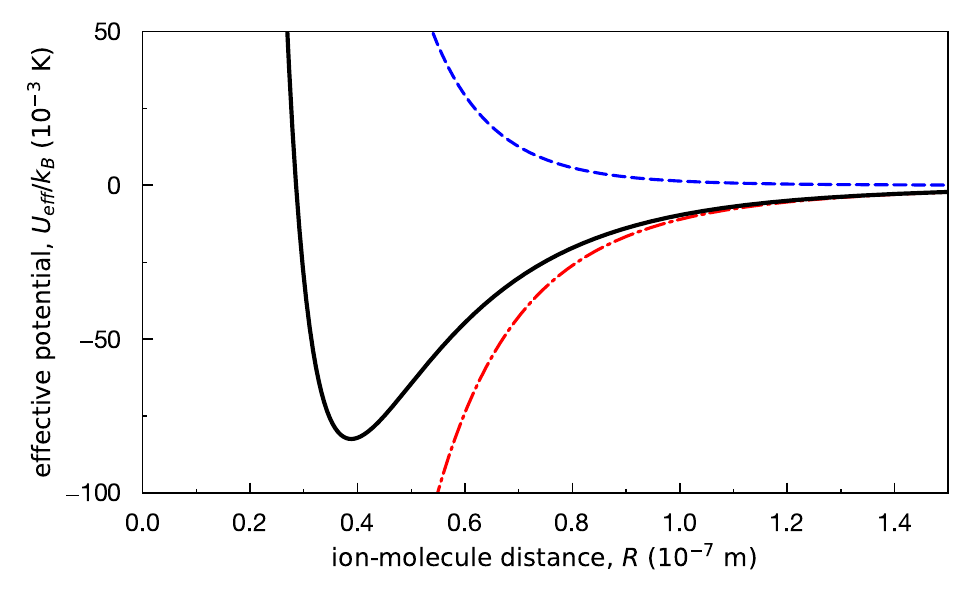}
		\end{center}
		\caption{\protect\rule{0ex}{0ex} Effective radial potential $U_{\textrm{eff}}/k_{\textrm{B}}$ per particle between a $ ^{138} $Ba$^{+} $ ion and a quantum gas of $^{23} $Na$ ^{39} $K molecules in their rovibronic ground state $ \ket{\textit{X} ^{1}\Sigma^{+}, v = 0, J = 0} $ as a function of the ion-molecule distance $R$. We assume a moderate number density of $ \textit{n} \approx 4 \times 10^{10} ~ \textrm{cm}^{-3} $. The balance between attractive ion-neutral (charge-dipole, $ \propto R^{-2} $, red dashed-dotted line) and repulsive molecule-molecule (dipole-dipole, $ \propto R^{-3} $, blue dashed line) interactions gives rise to a global minimum at $R_{\textrm{min}}$.
		}
		\label{fig:BaNaKcompetingInteractions}
	\end{figure}

	As a potential scenario where the competition between attractive and repulsive interactions would play an important role we consider an ensemble of homogeneously distributed molecules at a distance $R$ around the ion. In this case, while the molecules approach the ion, the collective dipole-dipole term which is proportional to $d_0^2/R_{\textrm{ij}}^3$ will dominate at short range. For the sake of simplicity, we have neglected the short-range van der Waals interaction $U_{\textrm{vdW}} = -C_6/R_{\textrm{ij}}^6$, and consider only next-neighbor contributions between the molecules. We further take into account the finite induced electric dipole moment $d(\mathcal{E}\left(R\right) )$ of the molecules (as shown in Fig.~\ref{fig:polarization}) and the finite projection of $ \vec{d}_{\textrm{i}}$ onto $ \vec{d}_{\textrm{j}}$. The intermolecular distance is $R_{\textrm{ij}} = 2 R \sin( \sqrt{\pi / N})$ and the effective dipole moment is $d_{\textrm{eff}}(R) = d(R) ( 1 - \sin( \sqrt{\pi/N} )^2 )$. Under these conditions and for an initial density of $ n \approx 4 \times 10^{10} ~ \textrm{cm}^{-3} $ we obtain the effective radial potential $U_{\textrm{eff}}$ for a single molecule depicted in Fig.~\ref{fig:BaNaKcompetingInteractions}. We find that the competition of repulsive intermolecular and attractive ion-molecule interactions in the considered configuration gives rise to a local minimum around $ R_{\textrm{min}} \approx 40 ~ \textrm{nm} $ which is shifted towards larger $R$ for an increasing number of molecules while its depth $U_{\textrm{eff}}/k_{\textrm{B}} \approx 80 ~ \textrm{mK} $ decreases for larger $N$.

	For simplicity, we have focused on an electrostatic view and omitted dynamical aspects such as a mechanism to dissipate kinetic energy once $R_{\textrm{min}}$ has been reached or a detailed stability analysis, which would be important for future realistic theoretical or experimental studies of such shell-like configurations in a 3D bulk gas. A potentially more accessible approach to studying the interplay of several simultaneously active interactions and their impact on the dynamics of a many-body system would be to immerse an ion into an ensemble of molecules confined to two dimensions. The latter can be realized, e.g. by utilizing 1D optical standing waves as a periodic trapping potential. In such setups the trapped molecules arrange themselves in layers with a regular spacing typically on the order of the wavelength of the dipole trap $\lambda$. This ensures very high trapping frequencies along the axis perpendicular to the trap layers such that the motional degrees of freedom are effectively constrained to a quasi-2D geometry. For an ion positioned at the center of a lattice site, that is for an ion-lattice spacing of $d_{\textrm{lat}}=0$, we calculate the effective potential of a molecule within the ensemble. In this geometry, we obtain $R_{\textrm{ij}} = 2 R \sin( \pi / N)$ and $d_{\textrm{eff}}(R) = d(R) \cos( \pi/N )$. Under such conditions, we again find local minima in the effective potential for a wide range of initial molecular densities.
	
	Such effects are an interesting field for future investigations, providing connections to polaron-type systems \cite{Goold2010,Schurer2017,Astrakharchik2023,Christensen2022,Chowdhury2024} and mesoscopic molecular ions \cite{Cote2002} which have been studied extensively in the context of ion impurities in an atomic gas. One possible direction would be to explore potential transitions between, e.g., superfluid, Wigner crystal \cite{Buechler2007,Astrakharchik2007} or supersolid phases \cite{Schmidt2022} induced by embedding ions in dense polar quantum gases. For instance, the predicted quantum phase transition to a crystalline phase with a triangular lattice structure in 2D bosonic dipolar gases requires a dimensionless density of $n_{\textrm{2D}} a_{\textrm{dd}}^2 > 290 $ \cite{Astrakharchik2007}. While this is out of reach in existing contemporary experiments using polar molecules in 1D optical lattices, e.g. fermionic KRb \cite{Valtolina2020}, NaK has been suggested as a suitable molecule due to its larger PDM \cite{Matveeva2012}. Under the assumption that, e.g. $N=500$ molecules can be arranged in a stable configuration as the one discussed above in a realistic experiment, the system would yield a dimensionless density of $n_{\textrm{2D}} a_{\textrm{dd}}^2 \approx 3 \times 10^4 $, which would make it promising for observing such a phase transition.
	
	\subsection{Ion-mediated shielding of reactive and inelastic collisions}\label{ion_shielding}
	In the following, we consider the impact of high molecular densities expected for the configurations discussed in the previous Section assuming they may be realized. For the example above, typical separations between NaK molecules would be $R_{\textrm{ij}} \approx 10 ~ \textrm{nm}$. This is comparable to the characteristic van der Waals length of NaK $\bar{a} \approx 10 ~ \textrm{nm}$ \cite{Julienne2011} and corresponds to strongly increased local densities approaching $n \approx 10^{18} ~ \textrm{cm}^{-3}$. In atom-ion mixtures, such high densities are inaccessible so far due to the extremely short lifetimes resulting from enhanced TBR \cite{Haerter2012,Krukow2016,Dieterle2020,Pandey2024} and resonantly enhanced multi-photon ionization (REMPI) \cite{Haerter2013}. Ensembles of neutral molecules are also known to exhibit two- and three-body losses with rates close to the universal limit attributed to sticky short-range collisions \cite{Mayle2013,Christianen2019,Croft2023,Jachymski2022}. Since collisional shielding techniques have not been demonstrated for densities in this region, we expect that the molecular ensembles would not be stable without a substantial polarization and the quasiplanar topology of the system.
	
	In the present scenario, the increase of local density is caused by the interaction of the molecules with the electric field of the ion. Due to their increasing alignment in the electric field mutual dipole-dipole repulsion eventually dominates at short range, which would result in the local minimum of the effective potential shown in Fig. \ref{fig:BaNaKcompetingInteractions} corresponding to a tightly confining shell. Due to the topologic similarity of a spherical surface with a planar disk (the corresponding spaces are homeomorphic up to a single point on the sphere), this confinement to a surface in 3D can be mapped to an equivalent geometry of polar molecules confined in a 2D plane, e.g., in a single site of a 1D optical standing wave. Since the radial electric field of the ion is locally parallel to the normal of any surface element on the shell, the configuration can be described by a homogeneous external electric field perpendicular to the plane containing molecules with individual effective dipole moments aligned parallel to the field and to each other. This situation where head-to-tail collisions are suppressed has been extensively studied with polar molecules, where efficient collisional shielding has been observed in degenerate Fermi gases of KRb \cite{Valtolina2020,Matsuda2020}. With the large degree of dipole moment alignment present in the discussed configuration, we therefore expect highly efficient shielding of two and three-body processes. This should result in a suppression of sticky collisions, TBR losses and REMPI. As all molecules on the shell would be separated from the ion, we also expect that ion-neutral TBR would be efficiently suppressed. Notably, this ion-mediated shielding mechanism would persist for or even benefit from very high local densities which are a prerequisite for entering the deep quantum-dominated regime of interactions. Similarly, we also expect efficient ion-mediated shielding for the case of dense molecular gases in a quasi-2D geometry achieved by confinement in a 1D optical lattice as discussed above.

	\section{Applications of hybrid ion-molecule systems}\label{applications}
	\subsection{Control of collisional processes in molecular gases}
	Applying sufficiently large homogeneous electric fields along the strongly confining axis has been shown to suppress short-range collisions by preventing head-to-tail collisions \cite{Valtolina2020}. In the proposed ion-molecules experiments a similar arrangement can be realized by replacing the external electric field with a single ion positioned between two lattice sites. The capability to control the ion-lattice site distance $ d_{\textrm{lat}} $ with the sub-wavelength precision necessary for avoiding direct contact between the ion and the molecules was demonstrated in several early works \cite{Guthoehrlein2001,Karpa2013}. For a common wavelength of a NIR dipole trap of $\lambda = 1064 ~\textrm{nm}$, the electric field of an ion displaced by $\lambda/4$, that is, 0.5 of the lowest lattice site spacing, would lead to an equally small induced dipole moment of NaK in two adjacent sites of $0.03 ~d_0$. Considering the van der Waals potential between two molecules, the effective barrier stemming from induced dipole-dipole repulsion is then $U_0 / k_{\textrm{B}} = 234 ~\textrm{nK} \approx T_{\textrm{mol}} $. This is on the order of the temperature in typical non-degenerate molecular quantum gases such that no significant change of the inelastic collision rate is expected. However, moving the ion closer to a lattice site gives rise to a substantial barrier. As an example, at an ion-lattice site spacing of $d_{\textrm{lat}} = ( \lambda/8, ~ \lambda/16, ~ \lambda/32) \approx (133, ~ 66, ~ 33) ~\textrm{nm}$ the barrier height increases to $U_0 / k_{\textrm{B}} = (0.06 , ~ 6.2, ~ 52.1) ~\textrm{mK}  \gg T_{\textrm{mol}}$. This suggests that ion-mediated shielding of short-range collisions may also be achieved using an ion as a nanoscale source of external electric fields without immersing it into the molecular cloud as discussed in Section \ref{ion_shielding}. An additional advantage of this approach is that changing $d_{\textrm{lat}}$ simultaneously lowers the barrier in the adjacent lattice site, e.g. to $U_0 / k_{\textrm{B}} <  10 ~\textrm{nK} \ll T_{\textrm{mol}}$ for $d_{\textrm{lat}} = \lambda/4$ to $\lambda/8$. In this fashion, systematic effects between different realizations of an experiment can be reduced allowing for precise differential measurements of the influence of electric fields on the dynamical properties of molecular ensembles such as the rates for elastic and reactive or sticky collisions.
	
	\subsection{Application to quantum technologies}
	One way to harness the advantage of quantum computation already with the limited computational resources of platforms available today, is to map specific problems onto simplified models that can be implemented on digital quantum simulators \cite{Daley2022}. For instance, a detailed understanding of solubility processes on the quantum level is of high practical importance, e.g. for the development of drugs. In this context, ions immersed into molecular quantum gases may exhibit properties similar to atomic cations embedded in a dense polarizable medium (e.g., liquid or solid), a commonly encountered system.  To this end, recent theoretical studies in ion-atom systems suggest that neutral media with a suitable polarizability may be used for simulations of solubility due to a strong enhancement of interactions between ions and neutrals, as well as between the solvent particles, even though the involved densities are low compared to those of typical liquids \cite{Chowdhury2024}. Similarly, the additional enhancement of the polarizability expected in the ion-molecule system discussed in this work may allow highly precise analog quantum simulations of more complex solubility processes that are currently studied using digital quantum simulators \cite{Castaldo2022}.

From the perspective of quantum information processing, atomic ions are among the most advanced high-performance platforms in the field \cite{Leibfried2003}. Complementary to these developments, the tremendous progress achieved in the control over ultracold molecules has also brought forward promising protocols for realizing gate operations by exploiting dipole-dipole interactions \cite{DeMille2002}, or by making use of the internal rotational states and their coupling to microwave fields to implement and manipulate qudits \cite{Sawant2020}. More recent ideas involve Rydberg atoms to establish and distribute entanglement between molecules \cite{Zhang2022,Wang2022}. In addition to these prominent examples, hybrid approaches promise to combine the most advantageous properties of different experimental systems. Along these lines, the combination of ions and polar molecules may prove a promising extension to the quantum toolbox of platforms using ions, atoms or molecules. As an example, in comparison to quantum computation protocols based on the interaction between ions and neutral atoms \cite{Doerk2010}, a hybrid approach motivated by early experiments on ion-atom collisions, the long-range interaction between an ion and polar molecules is enhanced by the ratio of their respective static polarizabilities, that is by a factor of typically $10^2$ to $10^4$. For instance, at field strengths above $\mathcal{E}^{\star}$ corresponding to ion-neutral distances of less than $ 100 ~\mathrm{nm}$ the long-range interaction between Ba$^{+}$ and NaK is about 3000 times larger than the charge-induced dipole interaction with a Rb atom. In the future, these expected properties and the capability to exert forces that depend on the rotational state of a molecule or the electronic state of an ion may be investigated in view of potential applications to ion-molecule interfaces or hybrid approaches to quantum computation.
	\section{Conclusion}
	In summary, we have discussed the potential realization of a hybrid ion-molecule system suitable for studies of their interactions in the quantum dominated regime. We expect that the combination of currently available techniques and methodology from the fields of ion-atom interactions, in particular bichromatic dipole traps and optimized rf-based hybrid apparatus as well as protocols for the generation of ground-state polar molecules, is feasible within the scope of the envisioned studies. We identify and discuss several promising research directions in different regimes of the molecular ensembles, trap configurations and geometries. Such investigations include reactive inter-species processes with several open channels for chemical reactions and the creation of polyatomic molecular ions. Based on simple hypotheses, we expect that despite the seemingly much higher complexity compared to ion-atom or neutral molecule systems, the additional degrees of freedom, foremost the rotational states of the molecules, may be exploited and controlled to an extent allowing for the realization of molecule-ion collisions in the ultracold regime, as well as sympathetic cooling. Furthermore, we found indications that the interplay between attractive and repulsive interactions induced by the presence of the ion in a dense polar medium may play an important role in the dynamics of the many-body system.
	
	These capabilities open the way to access the sought-after regime of quantum interactions between ions and neutrals holding great promise to advance several fields and applications including refined control of collisional properties in ensembles of polar molecules, quantum chemistry, and quantum-many-body physics, as well as quantum simulations and technologies.
	
	\section*{Acknowledgments}
L.K. thanks the DFG (German Research Foundation) for support through the Heisenberg Programme No. 506287139, and Germany’s Excellence Strategy EXC-2123 Quantum-Frontiers No. 390837967. We acknowledge financial support through the ANR/DFG project OpEn-MInt (ANR-22-CE92-0069-01). We are indebted to Mirco Siercke and Silke Ospelkaus for stimulating and insightful discussions. Romain Vexiau is gratefully acknowledged for providing us preliminary data on various diatomic molecules prior to publication.

\end{document}